\documentstyle[12pt]{article}
\begin{document}

\title{Quantum-Classical System:\\
 Simple Harmonic Oscillator}
\author{Tri Sulistiono\dag \\
Department of Physics, Institute of Technology Bandung\\
Jalan Ganesha 10, Bandung 40132, Indonesia}
\maketitle
\begin{abstract}
Problems concerning with application of quantum rules on classical phenomena have been widely studied, for which lifted up the idea about quantization and uncertainty principle. 
Energy quantization on classical example of simple harmonic oscillator has been reviewed in this 
paper.
\end{abstract}

\section{Introduction}

In the past few years applications of quantum rules on many classical problems have been widely 
studied, both theoretically and experimentally\lbrack 1-3\rbrack. Such applications lifted up 
problems concerning with quantization and the uncertainty principle, which are not considered in 
the classical scheme. These treatments are of considerable importance these days owing to their 
prospective applications or even more establish a new field, for example, in quantum computation 
and quantum cryptography\lbrack 4-6\rbrack.

How can we realize classical problems in quantum scheme? Let us consider a simple one-dimensional 
classical harmonic oscillator of mass $m$ with kinetic energy $p^2/2m$ and potential energy 
$kq^2/2=m\omega^2 q^2/2$, where $k$ is a constant. Thus, it leads to the corresponding Hamiltonian

\bigskip

$$H=-\frac{\hbar^2}{2m} \frac{\partial^2}{\partial q^2}+\frac{1}{2} kq^2 \;\;\;\;\;\;\;\ldots (1).$$

\bigskip
\noindent
This describes a force $-kq$, for which Newton's second law

\bigskip

$$m\frac{d^2q}{dt^2}=-kx \;\;\;\;\;\;\;\ldots (2).$$

\bigskip
\noindent
has oscillating solutions $A\sin(\omega t)+B\cos(\omega t)$, with $\omega 
=(k/m)^{1/2}$ being the angular frequency of vibration, which allow us to rewrite the Hamiltonian into the form

\bigskip

$$H=-\frac{\hbar^2}{2m} \frac{\partial^2}{\partial q^2}+\frac{1}{2} m\omega^2 q^2 \;\;\;\;\;\;\;\ldots (3).$$

\bigskip

In the quantum case we introduce the equation

\bigskip

$$-\frac{\hbar^2}{2m} \frac{d^2u(q)}{dq^2}+ V(q)u(q)=Eu(q)\;\;\;\;\;\;\;\ldots (4).$$

\bigskip
\noindent
Equation (4) known as time-independent Schr\"{o}dinger equation and the solutions $u(q)$ 
are called wave function which always take the form 

\bigskip

$$\psi(q,t)=u(q){\rm exp}(-iEt/\hbar)\;\;\;\;\;\;\;\ldots (5).$$

\bigskip

\section{Quantum Simple Harmonic Oscillator}
\subsection{The energy quantization}

Let us use the quantum theory to the simple one-dimensional classical harmonic oscillator we have 
introduced above. In the quantum case we substitute the potential energy equation $kq^2/2$ into 
Eq. (4) to obtain

\bigskip

$$\frac{d^2u}{dz^2}+(2\epsilon-z^2)u=0\;\;\;\;\;\;\;\ldots (6).$$

\bigskip
\noindent
where

\bigskip

$$z=(\frac{m\omega}{\hbar})^{1/2}q \:\;\;\;\;\;{\rm and}\:\;\;\;\;\;\epsilon=\frac{E}{\hbar\omega}\:\;\;\;\;\;\;\ldots (7).$$

\bigskip

As ussual, one expects solutions of Eq. (6) to show rapid decline as 
$z\rightarrow\pm\infty$. Our inspection of the asymptotic form, i.e. $z^2\gg\epsilon$, show 
$u\sim{\rm exp}(-z^2/2)$ is a solution in this region. This therefore suggest general solutions 
of the form $F(z){\rm exp}(-z^2/2)$, where $F$ is a polynomial. Substituting this form into Eq. (6) yields

\bigskip

$$\frac{d^2F}{dz^2}-2z\frac{dF}{dz}+(2\epsilon-1)F=0\:\;\;\;\;\;\;\ldots (8).$$
	
\bigskip

Suppose the leading term of $F$ is $z^n$. This contributes

\bigskip

$$n(n-1)z^{n-2}-2nz^n+(2\epsilon-1)z^n\;\;\;\;\;\;\;\ldots (9).$$

\bigskip
\noindent
to the left-hand side of Eq. (8). The coefficient of $z^n$ must vanish to comply with 
Eq. (8) and as lower-order terms in the polynomial $F$ only contribute to $z^{n-1}$, or 
lower powers, we demand from Eq. (9) that

\bigskip

$$\epsilon=n+\frac{1}{2} \;\;\;\;\;\;\;\;\;\;\; n=1,2,3,\ldots \:\;\;\;\;\;\;\ldots (10).$$

\bigskip

It follows from Eq. (7) that the energy $E$ is restricted to discrete levels given 
by

\bigskip

$$E_n=(n+\frac{1}{2})\hbar\omega \;\;\;\;\;\;\;\ldots (11).$$

\bigskip
\noindent
These levels have the interesting property that they are equispaced and the classical frequency 
$\omega$ is related to $E$ in the same way as the photon relation, $E={\it h}\nu$. This 
is no coincidence.

We have showed that the energy of quantum oscillator is quantized. The ground state, however, has 
energy $\frac{1}{2} \hbar\omega$ which, as in previous example, is above the classical minimum ($E=0$). The ground state wave function $u_0$ is given by $n=0$ in which case $F$ is constant, 
so $u_0\propto{\rm exp}(-z^2/2)$. Applying the normalization condition gives a gaussian function

\bigskip

$$u_0=(m\omega/\pi\hbar)^{1/4}{\rm exp}(-m\omega q^2/2\hbar)\:\;\;\;\;\;\;\ldots (12).$$

\bigskip

The expectation values of $q$ and $V$ for the ground state are

\bigskip

$$\langle q\rangle=\int_{-\infty}^\infty P_0(q)q\,dq =\int_{-\infty}^\infty u_0^2q\,dq =0 \:\;\;\;\;\;\;\;\ldots (13).$$

\bigskip

$$\langle V\rangle=\frac{k}{2}\int_{-\infty}^\infty P_0(q)q^2\,dq =\frac{k}{2}\int_{-\infty}^\infty u_0^2q^2\,dq =\frac{1}{2}E_0\:\;\;\;\;\;\;\;\ldots (14).$$

\bigskip
\noindent
respectively

\subsection{Further disscusion on quantization}

We can see that the Hamiltonian formulation in Eq. (1) take the form of the left-hand side of Eq. (4) by cancelled the $u(q)$ and inserting $V=\frac{1}{2}kq^2$.
By putting the wave function $\psi$ instead of $u(q)$ and write $E$ as $i\hbar\frac{\partial}{\partial t}$ we have the Schr\"{o}dinger equation in the form

\bigskip

$$-\frac{\hbar^2}{2m} \frac{d^2\psi}{dq^2}+ \frac{1}{2}m\omega^2q^2\psi=i\hbar \frac{\partial\psi}{\partial t}\;\;\;\;\;\;\;\ldots (15).$$

\bigskip
\noindent
By scalling the oscillator by introducing $s=q/q_0$ and $q_0^2=\hbar/m\omega$ yields the Schr\"{o}dinger equation in terms of $s$ for the wave function $\psi (s,t)$

\bigskip

$$i\hbar\frac{\partial\psi}{\partial t}=\hbar\omega (-\frac{1}{2}\frac{\partial^2\psi}{\partial s^2}+\frac{1}{2}s^2\psi )\;\;\;\;\;\;\;\;\ldots (16).$$

\bigskip
\noindent
The $\hbar$ may now be cancelled remaining only one parameter $\omega$ to define the oscillator. This scalling allow us to find the eigenfunctions for stationary states 
which satisfy

\bigskip

$$E\psi=\hbar\omega (-\frac{1}{2}\frac{\partial^2\psi}{\partial s^2}+\frac{1}{2}s^2\psi )\;\;\;\;\;\;\;\;\;\ldots (17).$$

\bigskip

We have already seen that the spectrum is discrete (Eq. (11)), and that the first few eigenfuctions all contain a factor ${\rm exp}(-\frac{1}{2}s^2)$ and the other 
factor being a polynomial. We can setting

\bigskip

$$\psi=f(s){\rm exp}(-\frac{1}{2}s^2)\;\;\;\;\;\;\;\;\;\;\;\;\ldots (18).$$

\bigskip
\noindent
by differentiation we have

\bigskip

$$\psi''=(f''-2sf'+(s^2-1)f){\rm exp}(-\frac{1}{2}s^2)\;\;\;\;\;\;\;\;\;\;\ldots (19).$$

\bigskip
\noindent
Inserting in the original equation now gives the equation for f,

\bigskip

$$Ef=\hbar\omega(-\frac{1}{2}f''+sf'+\frac{1}{2}f)\;\;\;\;\;\;\;\;\;\;\;\;\ldots (20).$$

\bigskip

We may now solve the problem using the power series approach. We see that none of terms is singular for any value for $s$, 
so we may expand $f$ as a Taylor series which convergent for all $s$,

\bigskip

$$f(s)=\sum_{j=0}^\infty c_j s^j \;\;\;\;\;\;\;\;\;\;\;\;\ldots (21).$$

\bigskip
\noindent
Substitute in the equation and collecting powers of $s^j$ in the result gives

\bigskip

$$Ec_j=\hbar\omega\{-\frac{1}{2}(j+2)(j+1)c_{j+2}+sc_j+\frac{1}{2}c_j\} \;\;\;\;\;\;\;\; {\rm for}j\geq 0 \;\;\;\;\;\;\;\;\;\;\ldots (22).$$

\bigskip
\noindent
that is,

\bigskip

$$\frac{1}{2}\hbar\omega(j+2)(j+1)c_{j+2}=\{ (j+\frac{1}{2})\hbar\omega-E\}c_j \;\;\;\;\;\;\;\; (j\geq 0)\;\;\;\;\;\;\;\;\;\;\;\;\ldots (23).$$

\bigskip
\noindent
We see that if we know $c_0$ and $c_1$, any subsequaent coefficient may be obtained by applying this relation a sufficient 
number of times. It remains ensure that $\psi (s)\rightarrow 0$ as $|s|\rightarrow\infty$.

Now we have for large $j$

\bigskip

$$\frac{c_{j+2}}{c_j}=\frac{2}{j}+0(\frac{1}{j^2}) \;\;\;\;\;\;\;\;\;\;\;\;\;\ldots (24).$$

\bigskip
\noindent
we can see that the coefficients show a behaviour similar to those in the Taylor series of exp $s^2$. 
Consequently, $\psi(\equiv f\, {\rm exp}(-\frac{1}{2}s^2))$ is inevitably unbounded as $s$ goes to infinity in either direction, 
and this will not do. The series for $f$ must therefore terminate and this happens only if 

\bigskip

$$E=E_n\equiv (n+\frac{1}{2})\hbar\omega \;\;\;\;\;\;\;\;\;\;\; (n\, {\rm interger} \geq 0) \;\;\;\;\;\;\;\;\;\;\ldots (25).$$

\bigskip
\noindent
when the final term in series will be $c_ns^n$. 
We can compute first few unnormalized function including their time dependence as given bellow,

\bigskip

$$\psi_0={\rm exp}(-\frac{1}{2}s^2)\,{\rm exp}(-\frac{1}{2}i\omega t).$$
$$\psi_1=s\, {\rm exp}(-\frac{1}{2}s^2)\,{\rm exp}(-\frac{3}{2}i\omega t).$$
$$\psi_2=(2s^2-1)\,{\rm exp}(-\frac{1}{2}s^2)\,{\rm exp}(-\frac{5}{2}i\omega t) \;\;\;\;\;\;\;\;\;\;\;\ldots (26).$$

\bigskip
\noindent
respectively

\subsection{The algebraic formulation of the simple harmonic oscilator}

We shall carry out the solution of the simple harmonic oscillator using an algebraic formalism based on the Dirac notation. 
The starting point is the fact that the Hamiltonian is almost factorizable as the difference of two squares,

\bigskip

$$(s-\frac{d}{ds})(s+\frac{d}{ds})\psi(s)=(s^2-\frac{d^2}{ds^2}-1)\psi(s)\;\;\;\;\;\;\;\;\;\;\;\ldots (27).$$

\bigskip

We shall introduce the s-representation by writing $\psi(s)=\langle s|\psi\rangle$ and define two operators ${\sf a}, {\sf a}^+$ 
by the relations

\bigskip

$$\langle s|{\sf a}=2^{-1/2}(s+\frac{d}{ds})\langle s| \;\;\;\;\;\;\;\;\;\;\;\;\ldots (28).$$

\bigskip

$$\langle s|{\sf a}^+=2^{-1/2}(s-\frac{d}{ds})\langle s| \;\;\;\;\;\;\;\;\;\;\;\;\ldots (29).$$

\bigskip
\noindent
Then

\bigskip

$$\hbar\omega\langle s|{\sf a}^+{\sf a}=\frac{\hbar\omega}{2}(s+\frac{d}{ds})(s-\frac{d}{ds})\langle s|=
\langle s|({\sf H}-\frac{1}{2}\hbar\omega) \;\;\;\;\;\ldots (30).$$

\bigskip
\noindent
whence

\bigskip

$${\sf H}=\hbar\omega({\sf a}^+{\sf a}+\frac{1}{2})=\hbar\omega(n+\frac{1}{2}) \;\;\;\;\;\;\;\;\ldots (31).$$

\bigskip
\noindent
which show us about the quantization on the energy of the oscillator.

\subsection{The Wilson-Sommerfeld rules of quantization}

This rules discovered by W. Wilson and A. Sommerfeld independently in 1915\lbrack 7\rbrack. This method consists in solving 
the classical equation of motion in the Hamiltonian form, therefore applying the coordinates $q_1, \ldots , q_{3n}$ 
and the canonically conjugate momenta $p_1, \ldots , p_{3n}$ as the independent variables. 
The assumption is the introduced that only those classical orbits are allowed as stasionary states for which the 
following condition are satisfied,

\bigskip

$$\oint p_k\, dq_k=n_k h, \;\;\;\;\; k=1,2,\ldots ,3n \;\;\;\;\;\; n_k={\rm an\, integer} \;\;\;\;\;\;\;\;\ldots (32).$$

\bigskip

This integrals can be calculated only for conditionally periodic systems, i.e. for systems for which coordinates 
can be found each of which goes through a cycle as a function of the time, independently of the others. Sometimes 
the coordinates can be chosen in several different ways, in which case the shapes of the quantized orbits depend 
on the choice of coordinate systems, but the energy values do not.

\section{Concluding Remark}

It follows from Eq. (14) and the followed description that the expectation value of the kinetic energy 
$\langle E_0\rangle-\langle V\rangle$ is $\frac{1}{2}E_0$ also. As in the classical case, the 
average kinetic and potential energies are the same. This remains true for the excited levels ($n\geq 1$). We 
could construct such levels wave function by substitute a full polynomial for $F$ in Eq. 
(8) and equate the coefficients of all the powers (not just $z^n$) to zero.

\subsection*{Acknowledgements}
We wish to acknowlegde the kindness of Professor P.C.V. Davies and J. L. Martin for publishing {\it Quantum 
Mechanics}\lbrack 8\rbrack and {\it Basic Quantum Mechanis}\lbrack 9\rbrack respectively, for which most of materials in this paper related to.

\subsection*{References}
\begin{flushleft}
\dag e-mail address: rezan@melsa.net.id\\
\lbrack 1\rbrack R.F. Fox and B.L. Lan, Phys. Rev. A {\bf 41}, 2952 (1990).\\
\lbrack 2\rbrack B.L. Lan and R.F. Fox, Phys. Rev. A {\bf 43}, 646 (1991).\\
\lbrack 3\rbrack E.G. Harris, Phys. Rev. A {\bf 42}, 3685 (1990).\\
\lbrack 4\rbrack D. Deutsch, Proc. R. Soc. London A {\bf 400}, 997 (1985).\\
\lbrack 5\rbrack P. Shor, in {\it Proceedings of the 35th Annual Symposium on the Foundations of 
Computer Science} (IEE Computer Society, Los Alamitos, CA, 1994) p. 124.\\
\lbrack 6\rbrack A. Ekert and R. Jozsa, Rev. Mod. Phys. {\bf 68}, 733 (1985).\\
\lbrack 7\rbrack A. Sommerfeld, Ann. d. Phys. {\bf 51}, 1 (1916); W.Wilson, Phil. Mag. {\bf 29}, 795 (1915)\\
\lbrack 8\rbrack P.C.V. Davies, {\it Quantum Mechanics}(Routledge \& Kegan Paul, London, 1984).\\
\lbrack 9\rbrack J.L. Martin, {\it Basic Quantum Mechanics}(Clarendon Press, London, 1981).\\
\end{flushleft}
\end{document}